\documentclass[letterpaper,10pt]{article} 

\usepackage{osameet3} 
\usepackage{upgreek}

\usepackage{amsmath,amssymb}
\usepackage[colorlinks=true,bookmarks=false,citecolor=blue,urlcolor=blue]{hyperref} 

\begin{document}

\title{Toward integrated synchronously pumped optical parametric oscillators in silicon nitride}

\author{Niklas M. Lüpken$^{1,*}$, David Becker$^{1}$, Thomas Würthwein$^{1}$, Klaus-J. Boller$^{2}$, and Carsten Fallnich$^{1,2,3}$}

\address{$^1$Institute of Applied Physics, University of Münster, Corrensstraße 2, 48149 Münster, Germany \\
$^2$MESA+ Institute for Nanotechnology, University of Twente, P.O. Box 217, Enschede 7500 AE, The Netherlands \\
$^3$Cells in Motion Interfaculty Centre, University of Münster, Waldeyerstraße 15, 48149 Münster, Germany \\}
\email{*n.luepken@uni-muenster.de}

\copyrightyear{2021}

\begin{abstract}
We present a tunable, hybrid waveguide-fiber optical parametric oscillator (OPO) synchronously pumped by an ultra-fast fiber laser exploiting four-wave mixing (FWM) generated in silicon nitride waveguides. Parametric oscillation results in a 35\,dB enhancement of the idler spectral power density in comparison to spontaneous FWM, with the ability of wide wavelength tuning over 86\,nm in the O-band. Measurements of the oscillation threshold and the efficiency of the feedback loop reveal how an integration of the OPO on a single silicon nitride chip can be accomplished at standard repetition rates of pump lasers in the order of 100\,MHz.
\end{abstract}



\maketitle

\section{Introduction}
Optical parametric oscillators (OPOs) are versatile and widely tunable light sources of high interest for many applications, such as coherent Raman scattering microscopy \cite{Ganikhanov2006,Jurna2006}, generation of squeezed light \cite{Tanimura2006}, or mid-infrared spectroscopy \cite{Leindecker2011}. OPOs can provide pulses as short as tens of femtoseconds, average powers up to hundreds of milliwatts, and wavelength tunability, e.g. by dispersive tuning \cite{Klein2001, Yamashita2006, Brinkmann2016}, from 200\,nm to above 10\,$\upmu$m \cite{VanDriel1995}. However, existing synchronously pumped OPOs operating at around 100\,MHz repetition rates of standard pump lasers do not seem feasible for integration on a chip, because waveguide propagation losses are too high for realizing feedback resonators with several meters roundtrip length.

Chip-integrated OPOs with extremely high repetition rates, i.e., ranging from several THz down to several hundreds of GHz, have been realized as so-called Kerr comb generators \cite{Kippenberg2004, Stern2018, Gaeta2019}. Kerr combs find wide-ranging applications, e.g., for telecommunications \cite{Levy2012}, optical ranging \cite{Trocha2018}, dual-comb spectroscopy \cite{Dutt2018, Suh2016}, or optical clocks \cite{Papp2014}. However, the output pulse energies are low due to the high repetition rates, and wavelength tuning of the output, such as with dispersive tuning, would not work due to insufficient dispersive resonator length. Furthermore, because the energy is distributed over a broad spectral range, applications requiring narrow spectral bandwidths with high energies can not be driven by Kerr combs.

Accordingly at standard repetition rates, synchronously pumped OPOs have so far been realized mostly as fiber-based oscillators \cite{Sharping2008,Brinkmann2016,Yang2018,Brinkmann2019a}. However, this approach does not provide interferometric stability and does not allow for seamless embedding into photonic circuits and on-chip sensing strategies \cite{DeVos2007,Robinson2008a,Dhakal2014}. As an important step towards integration, waveguide-based OPOs have been demonstrated in a hybrid approach using optical fibers, enabling low repetition rates for synchronous pumping \cite{Kuyken2013,Wang2015i}. However, these demonstrations have put their focus solely on the low-bandgap material silicon for providing higher nonlinear gain or extending the spectral coverage \cite{Kuyken2013}. Specifically, the silicon waveguide platform does not appear suitable for on-chip integration with low repetition rates, high pulse energies, and high peak powers, because it suffers from significant linear as well as nonlinear losses which are difficult to compensate by nonlinear gain \cite{Liu2011}. 

Here, we demonstrate the first hybrid waveguide-fiber low-repetition rate OPO based on high-bandgap, low-loss silicon nitride (SiN) waveguides. The SiN platform is particularly promising for on-chip integration, because the reduced nonlinearity can potentially be more than compensated by dramatically lower loss \cite{Epping2015a, Roeloffzen2018} to enable an on-chip integration. However, using theoretical scaling arguments only is critical, because the achievable gain values strongly depend on specific assumptions regarding pulse duration, waveguide dispersion and parasitic loss. Furthermore, the damage threshold of the material has to be considered. Instead, a direct experimental determination of the central gain and roundtrip loss parameters is more promising and trustworthy. These parameters can be extracted from operating a hybrid OPO with a high-bandgap SiN waveguide as gain element, and an experimental characterization of the oscillation threshold as well as feedback efficiency as a function of the operation parameters. Using the experimental data of gain and loss of the hybrid OPO, hereinafter named waveguide-based OPO (WOPO), we calculate the lowest-possible repetition rate of a synchronously pumped all-chip SiN WOPO. As the result of scaling with experimental parameters, we show that repetition rates as low as approximately 66\,MHz are possible with chip-integration, such that established fiber-based pump sources could be used to drive the SiN-based WOPO.

\section{Experimental results}    \label{sec:experiments}


\subsection{Setup of the waveguide-based optical parametric oscillator}
\begin{figure}[htbp]
\centering\includegraphics[width=1\linewidth]{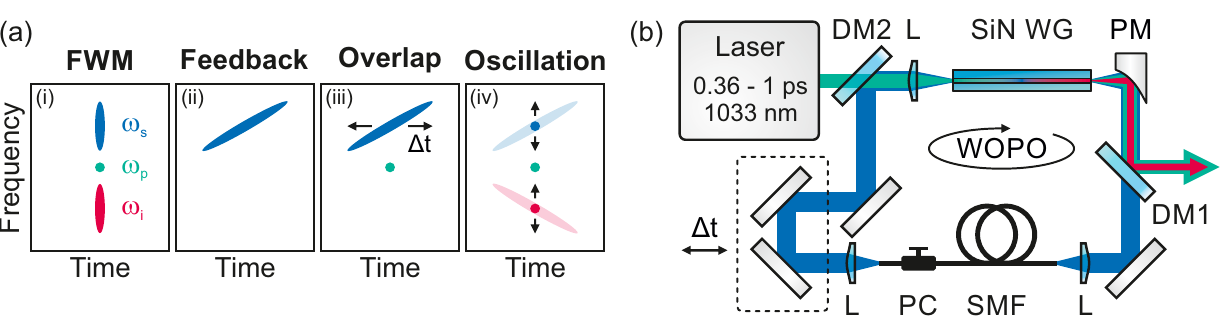}
\caption{(a) Schematic spectrograms of (i) spontaneous four-wave mixing (FWM) in the first roundtrip, (ii) the temporally stretched signal pulse, (iii) the adjustable temporal overlap of the stretched signal pulse with the subsequent pump pulses, and (iv) the stimulated FWM spectra, adjustable in frequency via the delay line. (b) Experimental setup of the WOPO. For details see text.}
\label{fig:setup}
\end{figure}

The working principle of the tunable WOPO is based on parametric frequency conversion by four-wave mixing (FWM, see spectrograms in Fig.~\ref{fig:setup}(a)): The first pump pulse at angular frequency $\omega_\text{p}$ generates broadband signal and idler sidebands by degenerate spontaneous FWM (case (i) in Fig.~\ref{fig:setup}(a)) at angular frequencies $\omega_\text{s}$ and $\omega_\text{i}$, respectively, obeying the energy conservation $2\omega_\text{p}=\omega_\text{s}+\omega_\text{i}$ as well as the phase-matching condition $2\beta_\text{p}=\beta_\text{s}+\beta_\text{i}$, where $\beta_j$ are the propagation constants \cite{Agrawal2013a}. One sideband, here the signal sideband, is fed back while the idler sideband and the residual pump are extracted. Wavelength tuning was realized by dispersive tuning as follows \cite{Klein2001, Yamashita2006, Brinkmann2016}. A dispersive element, here a single-mode fiber, stretches the signal pulse temporally by group velocity dispersion (case (ii)) and a delay ($\Delta t$) ensures temporal overlap with the subsequent pump pulses (case (iii)). The spectral part of the dispersed signal pulse that overlaps with the pump pulse acts as a seed to stimulate the FWM process, eventually resulting in parametric oscillation (case (iv)) when gain exceeds loss. The output frequency could be tuned by changing the delay and, therefore, overlapping a different spectral part of the signal pulse with the subsequent pump pulses.

In the experiments (see Fig.~\ref{fig:setup}(b)), the WOPO was pumped with an ytterbium-doped fiber laser (Satsuma, Amplitude Systemes) centered at 1033\,nm wavelength with a repetition rate of 1\,MHz. As the FWM process does not depend on the repetition rate, we denote pulse energy values instead of average power values in the following. In order to investigate the influence of the pump pulse parameters on the WOPO output, a Fourier filter (similar to Ref. \cite{Brinkmann2015}, but not shown) behind the pump laser was used to adjust the pump pulse duration between 0.36\,ps and 1.0\,ps. The pulse duration was set to 900\,fs and a combination of a half-wave plate and a polarizing beam-splitter (both not shown) was used to adjust the pump pulse energy for the WOPO. The pump pulses were coupled into a 7\,mm long, 950\,nm high, and 1350\,nm wide SiN waveguide (SiN WG, fabricated by LioniX International B.V.) embedded in silicon dioxide using an aspherical lens (L, $f=3.1\,\text{mm}$) in order to pump the FWM process. The input coupling efficiency of the pump pulses was determined by transmission measurements to about $-7\,\text{dB}$. The signal, idler, and residual pump beams were collimated with an off-axis parabolic mirror (PM) to avoid chromatic aberration. A subsequent dichroic mirror (DM1) reflected the residual pump and idler pulses and transmitted the signal pulses for feedback. Here, we coupled back the signal sideband as the idler output is of interest, e.g., for coherent Raman imaging \cite{Lupken2021a}, but in general the idler sideband could be coupled back alternatively to make use of the signal wave. Behind the dichroic mirror, the signal pulses were coupled into a 200\,m long single-mode fiber (SMF, group velocity dispersion coefficient $\beta_2=34\,\text{fs}^2/\text{mm}$ at 900\,nm signal wavelength) and spatially as well as temporally overlapped with the subsequent pump pulses via a dichroic mirror (DM2) for feedback. A polarization controller (PC) was used to align the polarization of the signal pulses with the polarization of the pump pulses for maximal feedback. The length of the feedback loop and, correspondingly, the output wavelength could be adjusted by means of dispersive tuning via a delay line ($\Delta t$).

\subsection{Output spectrum and wavelength tuning capability}

\begin{figure}[htbp]
\centering\includegraphics[width=1\linewidth]{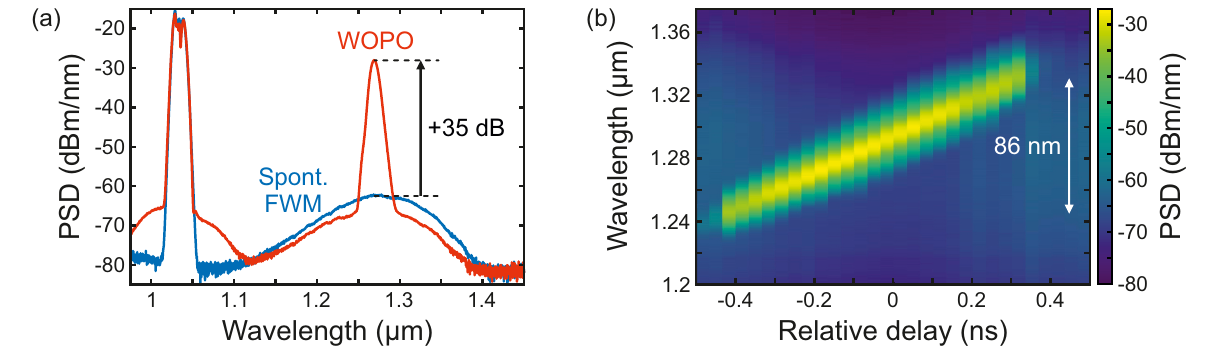}
\caption{(a) Power spectral density (PSD) of the WOPO output when the feedback was unblocked (red curve) and when the feedback was blocked (blue curve). (b) Color-coded idler spectra of the WOPO as a function of the delay of the signal pulses relative to the pump pulses.}
\label{fig:output}
\end{figure}

In order to characterize its output, the WOPO was pumped with pulses of 900\,fs pulse duration and 4\,nJ pulse energy. The pump pulses generated a broadband idler sideband by spontaneous FWM (see blue curve in Fig.~\ref{fig:output}(a)), when the feedback was blocked. The idler sideband spanned from about 1.1\,$\upmu$m wavelength to nearly 1.4\,$\upmu$m, matching numerical calculations of the parametric FWM gain \cite{Lupken2020a}. When the feedback was unblocked, the output spectrum of the WOPO (red curve) showed a 35\,dB enhanced power spectral density (PSD) of the idler sideband. The idler pulse energy of 28.5\,pJ corresponds to an external conversion efficiency of $-21.5$\,dB while the internal efficiency was estimated to about $-13$\,dB, which is higher than in silicon-based WOPOs with $-20.9$\,dB \cite{Wang2015i} or $-14.4$\,dB \cite{Kuyken2013}.

Via the delay $\Delta t$, the WOPO idler wavelength could be tuned over a wavelength span of 86\,nm from approximately 1246\,nm to 1332\,nm (see Fig.~\ref{fig:output}(b)) with a spectral bandwidth of about 10\,nm. The tuning range was limited by the parametric gain, which is represented by the spontaneous FWM spectrum (blue curve in Fig.~\ref{fig:output}(a)) and is determined by the dispersion profile of the waveguide. 

\subsection{Threshold and feedback}

In order to determine the necessary pump energy for a chip-integrated WOPO, the oscillation threshold was measured, and furthermore the minimum possible feedback and, therewith, the maximum tolerable resonator loss was retrieved. These data will be used in Sec.~\ref{sec:feasibility}, to estimate the feasibility of an integrated WOPO.

\begin{figure}[htbp]
\centering\includegraphics[width=1\linewidth]{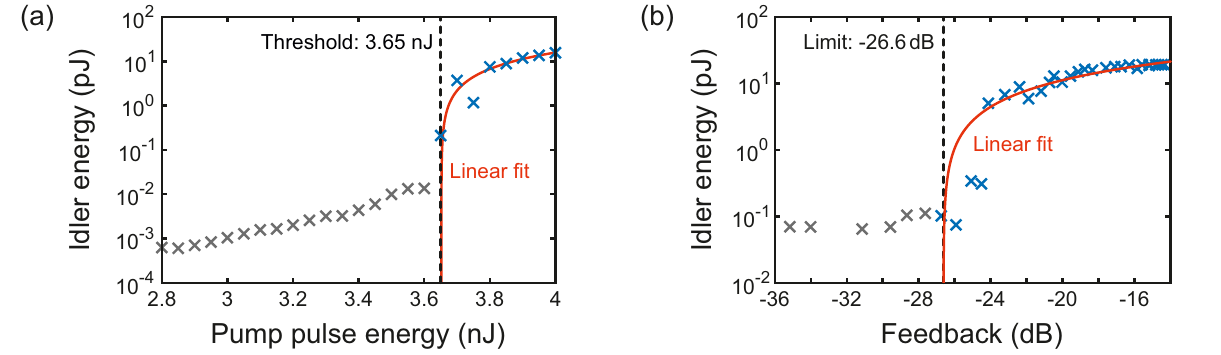}
\caption{(a) Measured idler energy (crosses) plotted logarithmically versus the pump pulse energy for a signal feedback of $-14\,$dB revealing an oscillation threshold of 3.65\,nJ (black dashed line), which was determined with a linear fit (red line) of the blue-colored data points. (b) Measured idler energy (crosses) plotted logarithmically versus the cavity feedback for a pump pulse energy of 4\,nJ, showing a lower limit of $-26.6\,\text{dB}$ (black dashed line) determined with a linear fit (red line) of the blue-colored data points.}
\label{fig:threshold}
\end{figure}

For the determination of the oscillation threshold, the idler energy was measured as a function of the pump pulse energy (see Fig.~\ref{fig:threshold}(a)). For these measurements, the efficiency of the feedback loop of the signal sideband was conservatively estimated to about $-14\,\text{dB}$ from a waveguide transmission of $-8.9\,\text{dB}$ and a transmission of the SMF of $-4.6\,\text{dB}$. A linear fit of the measured idler energy revealed an oscillation threshold of 3.65\,nJ, which is three orders of magnitude larger than for a WOPO based on silicon waveguides \cite{Wang2015i,Kuyken2013}. The higher threshold results from the three orders of magnitude lower nonlinearity of SiN compared to silicon. 
Near the oscillation threshold the presented WOPO showed some instabilities of its output (see measurement data between 3.65\,nJ and 3.75\,nJ in Fig.~\ref{fig:threshold}(a)) due to fluctuations of the input coupling and the feedback, however, this would be significantly improved by an on-chip integration of the WOPO resonator.

In order to explore the feasibility of an integrated WOPO, the experimentally required minimum feedback for oscillation was measured by reducing the feedback in a defined way by means of a variable neutral density filter in the feedback loop of the WOPO. With the pump pulse energy set to 4\,nJ the idler energy decreased linearly down to a feedback of $-26.6\,\text{dB}$ (blue crosses in Fig.~\ref{fig:threshold}(b)), which was estimated using a linear fit (red curve) applied to the blue-colored data points. At this feedback value the parametric oscillation stopped and only spontaneous FWM was measured (gray crosses). Hence, for stable parametric oscillation the necessary feedback had to be larger than about $-24\,\text{dB}$. With this result the overall losses are sufficiently low to enable an integration of the WOPO cavity on a single chip when taking the propagation losses of a SiN feedback loop into account, which will be discussed in Sec.~\ref{sec:feasibility}.

In future experiments, as the parametric amplification at threshold scales exponentially with the propagation length \cite{Agrawal2013a}, the presented results can be scaled-up by using longer waveguide lengths resulting in a reduced oscillation threshold and lower necessary feedback. For example, doubling the propagation length could reduce the oscillation threshold -- in a simple estimation -- by a factor of $\exp(2)=7.4$.

\subsection{Pump dependence}    \label{sec:pumpdependence}

Depending on the specific application, different output pulse parameters, such as spectral bandwidth or pulse duration, are required. However, the WOPO resonator does not allow for adjustments of the output pulse parameters, except for a wavelength tuning via changing the resonator length for dispersive tuning. Nevertheless, as the pump pulse parameters have shown to be crucial for the output pulse parameters within parametric amplification \cite{Lupken2021a}, we investigated the impact of the pump pulse duration on the oscillation threshold and the parametric amplification. Based on these investigations the optimal pump pulse parameters can be determined for the WOPO light source. 

\begin{figure}[htbp]
\centering\includegraphics[width=1\linewidth]{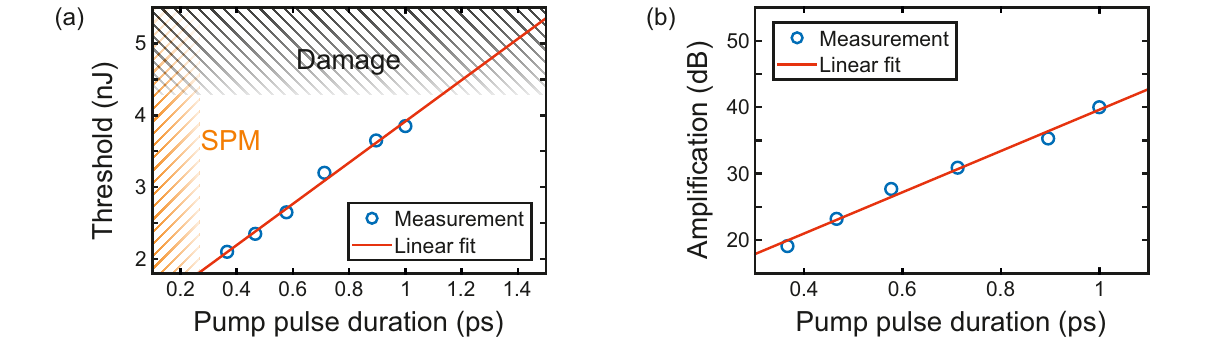}
\caption{(a) Oscillation threshold and (b) amplification of the power spectral density measured as a function of the pump pulse duration. The measurement data are depicted as blue circles and fits are shown as red lines. The black-shaded area depicts the region of the damage threshold and the orange-shaded area the region where the spectral evolution is dominated by self-phase modulation (SPM).}
\label{fig:pumpdependence}
\end{figure}

First, to explore the potential of reducing the oscillation threshold, the threshold was determined for different pump pulse durations using measurements of the idler output energy and a linear fit function as can be seen in Fig.~\ref{fig:threshold}(a). The oscillation threshold, plotted in Fig.~\ref{fig:pumpdependence}(a), was reduced to 2.1\,nJ for a pulse duration of 360\,fs, which was the shortest available pulse duration. The linear relationship between the oscillation threshold and the pump pulse duration originates from a higher necessary pulse energy for an increased pulse duration in order to reach the necessary peak power at the oscillation threshold. We used numerical simulations based on the nonlinear Schrödinger equation for pulse propagation according to Ref.~\cite{Poletti2008} to observe that the minimum pump pulse duration for the WOPO was limited to about 200\,fs, as then spectral broadening, induced by self-phase modulation (SPM), would dominate the spectral evolution for shorter pump pulses (orange-shaded area in Fig.~\ref{fig:pumpdependence}(a)). An oscillation threshold of 3.85\,nJ was measured for a pump pulse duration of 1\,ps. Even longer pulses could be used, but the maximum pump pulse duration was limited in our experiments by the increased oscillation threshold as well as the damage threshold \cite{Soong2011, Lupken2021a} of the waveguides (black-shaded area in Fig.~\ref{fig:pumpdependence}(a)). Therefore, only pump pulse durations between 200\,fs and 1\,ps were reasonable in the experiments to avoid SPM or damage of the waveguide.

Second, in order to determine the available parametric gain as a function of the pump pulse duration, the amplification, defined as the ratio between the power spectral densities of stimulated and spontaneous FWM (see Fig.~\ref{fig:output}(a)), was measured as a function of the pump pulse duration (shown in Fig.~\ref{fig:pumpdependence}(b)). The amplification reached 35\,dB for 0.9\,ps pump pulses, which confirms earlier measurements of optical parametric amplification experiments performed in SiN waveguides \cite{Kowligy2018,Lupken2021a} and PCFs \cite{Lefrancois2012}. An amplification of even 40\,dB was reached for 1\,ps pump pulses, however, instabilities of the output energy occurred due to the pump pulse energy being too close to the high oscillation threshold. Thus, the maximum amplification was limited by the increased oscillation threshold which increased with the pump pulse duration. 



\section{Feasibility of an integrated waveguide-based optical parametric oscillator}    \label{sec:feasibility}


The potential of an integration of the WOPO on a single chip is explored in this section. The feasibility of this integration is based on the measured properties shown in the previous section, where a SiN-based hybrid WOPO was demonstrated and investigated in detail on its operation parameters. The estimations are based on the pump parameters used in the experiments, which are currently not available on-chip but are representative for commercially available laser systems.


In order to derive a proper design for a fully integrated oscillator, the losses of the WOPO resonator have to be considered carefully: In an integrated WOPO, the resonator loss is dominated by the waveguide propagation loss, which can be minimized by reducing the resonator length and, thus, choosing higher repetition rates of the pump source. In the following, the minimum repetition rate of the pump source, which would enable an integration of the WOPO on a single chip, is determined by calculating the resonator loss for different resonator lengths (i.e. different pump repetition rates) and comparing these values with the measured necessary feedback for parametric oscillation (see Fig.~\ref{fig:threshold}(b)). 

\begin{figure}[htbp]
\centering\includegraphics[width=1\linewidth]{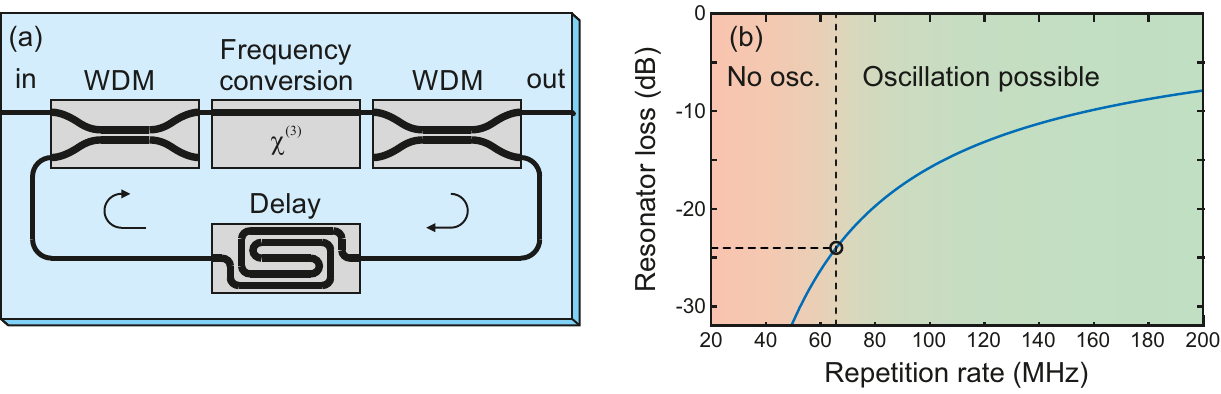}
\caption{(a) Schematic of an integrated WOPO. Heaters for dispersive tuning are not shown. (b) Resonator loss estimated for an integrated SiN WOPO as a function of the repetition rate. The SiN resonator loss reaches $-24$\,dB at around 66\,MHz (black dashed lines) from which on parametric oscillation should be possible (green-shaded area).}
\label{fig:reprate}
\end{figure}

An integrated WOPO has to include four essential components (see schematic in Fig.~\ref{fig:reprate}(a)): two wavelength-division multiplexers (WDMs), a frequency conversion section, and a dispersive delay closing the resonator. With such circuit design, the total WOPO resonator loss is the product of the insertion losses of the WDMs, the propagation losses in the frequency conversion, and those of the delay section. The loss in this delay section is a critical parameter in respect of its length and the accompanying loss, as the waveguide-based delay has to be longer than 1.6\,m (geometric length) for repetition rates in the order of 100\,MHz or below. The total propagation loss $\alpha_\text{prop} = L_\text{res}\eta$ depends on the resonator length $L_\text{res}$ and the propagation loss coefficient $\eta$ (in dB/m). In order to minimize the total propagation loss, the resonator length $L_\text{res}= c/(n_\text{eff}\, f_\text{rep})$ should be chosen as short as possible, which, in the case of synchronous pumping, depends on the typically fixed repetition rate of the pump source $f_\text{rep}$, the speed of light in vacuum $c$, and the effective refractive index $n_\text{eff}$. Therefore, a trade-off between lowering the repetition rate and decreasing the resonator loss has to be taken into account: On the one hand, for low repetition rates the resonator loss is too high to allow for parametric oscillation. On the other hand, for increased repetition rates the resonator loss decreases with the resonator length and, therefore, the propagation loss is reduced, eventually enabling parametric oscillation. However, sufficient peak power at high repetition rate is hardly achievable with conventional and compact mode-locked laser systems.

The lowest repetition rate, at which parametric oscillation would be possible, is calculated by comparing the resonator loss as a function of the repetition rate (see Fig.~\ref{fig:reprate}(b)) with the experimentally determined minimal required feedback loop efficiency of $-24\,\text{dB}$ (see Fig.~\ref{fig:threshold}(b)): The WDMs can be optimized such that their insertion losses become lower than 0.3\,dB \cite{Rao2020} and the propagation loss coefficient can be conservatively estimated as $\eta=0.1\,\text{dB/cm}$ \cite{Tien2010,Roeloffzen2018}. Assuming an effective index of 1.9, which was calculated with a mode solver (FIMMWAVE by Photon Design), the resonator loss will become lower than $-24$\,dB for a repetition rate of equal or higher than 66\,MHz. This should enable stable parametric oscillation due to a sufficiently strong feedback. 

Altogether, the above conservative estimation of the resonator loss clearly shows that an integrated WOPO on the basis of SiN should be possible, when pumped at a repetition rate in the order of 80 to 100\,MHz, thus, using conventional mode-locked (fiber) laser systems as the pump source.

Wavelength tuning of such WOPO can be accomplished by dispersive tuning on account of the resonator length in the order of meters. For instance, heaters can be integrated to change the resonator length sufficiently via the thermo-optic coefficient (generally $(2.5\pm0.5)\cdot 10^{-5}/\text{°C}$) \cite{Xue2016}, such that a relatively small temperature difference of $(37\pm7)\,\text{°C}$ of the WOPO resonator is already sufficient to accomplish the 86\,nm wavelength tuning demonstrated in Sec.~\ref{sec:experiments}. Alternatively, changing the wavelength or the repetition rate of the pump laser can be used for wavelength tuning \cite{Brinkmann2019a,Brinkmann2016}. Therefore, wavelength tuning by dispersive tuning is feasible for an integrated WOPO, which would not be possible with microring resonators.

Further improvements of the chip-integration of the WOPO can be accomplished with high-aspect ratio waveguides in the delay section to achieve ultra-low propagation losses in the order of $0.001\,\text{dB/cm}$ \cite{Bauters2011a,Puckett2021}. With this loss coefficient, repetition rates in the single-MHz range would be possible if according long resonator lengths can be realized on a chip, enabling amplified gain-switched diodes as pump sources. Moreover, the resonator loss could be further reduced by using the idler sideband for the feedback, in order to benefit from lower propagation loss due to the longer wavelength.

\section{Conclusion}

A synchronously pumped hybrid waveguide-fiber optical parametric oscillator (WOPO) in silicon nitride was demonstrated, showing a 35\,dB enhancement of the power spectral density in comparison to spontaneous four-wave mixing. The output wavelength of the WOPO could be tuned by more than 86\,nm via dispersive tuning. 

Measurements on the necessary feedback for parametric oscillation in a 7\,mm long waveguide with 4\,nJ of pump pulse energy revealed that $-24$\,dB feedback are sufficient for reliable parametric oscillation. A conservative estimation of the resonator loss showed that under these boundary conditions an integration of the parametric oscillator on a single chip should be possible for pump repetition rates larger than about 66\,MHz. Dispersive tuning can be applied to an integrated device via heaters in order to accomplish wavelength tuning. With the targeted miniaturization, for instance on-chip spectroscopic Raman experiments \cite{Lupken2020a,Lupken2021a} will become feasible.

\section*{Acknowledgments}
The authors thank Jörn P. Epping for support on the design and LioniX International B.V. for the fabrication of the silicon nitride waveguides.

\section*{Disclosures}
The authors declare no conflicts of interest.


\begin{thebibliography}{10}

\bibitem{Evans2008}
Conor~L. Evans and X.~Sunney Xie.
\newblock {Coherent Anti-Stokes Raman Scattering Microscopy: Chemical Imaging
  for Biology and Medicine}.
\newblock {\em Annu. Rev. Anal. Chem.}, 1(1):883--909, jul 2008.

\bibitem{Lotem1976}
Haim Lotem, R.~T. Lynch, and N.~Bloembergen.
\newblock {Interference between Raman resonances in four-wave difference
  mixing}.
\newblock {\em Phys. Rev. A}, 14(5):1748--1755, nov 1976.

\bibitem{Cheng2001a}
Ji-Xin Cheng, Andreas Volkmer, Lewis~D Book, and X~Sunney Xie.
\newblock {An Epi-Detected Coherent Anti-Stokes Raman Scattering (E-CARS)
  Microscope with High Spectral Resolution and High Sensitivity}.
\newblock {\em J. Phys. Chem. B}, 105(7):1277--1280, feb 2001.

\bibitem{Nestor1978}
J~R Nestor.
\newblock {Polarization properties of coherent anti-stokes Raman spectra (CARS)
  in isotropic liquids}.
\newblock {\em J. Raman Spectrosc.}, 7(2):90--95, apr 1978.

\bibitem{Oudar1979a}
Jean‐Louis Oudar, Robert~W. Smith, and Y.~R. Shen.
\newblock {Polarization‐sensitive coherent anti‐Stokes Raman spectroscopy}.
\newblock {\em Appl. Phys. Lett.}, 34(11):758--760, jun 1979.

\bibitem{Cheng2001}
Ji-Xin Cheng, Lewis~D Book, and X~Sunney Xie.
\newblock {Polarization coherent anti-Stokes Raman scattering microscopy}.
\newblock {\em Opt. Lett.}, 26(17):1341, sep 2001.

\bibitem{Wurthwein2017}
Thomas W{\"{u}}rthwein, Maximilian Brinkmann, Tim Hellwig, and Carsten
  Fallnich.
\newblock {Rapid spectro-polarimetry to probe molecular symmetry in multiplex
  coherent anti-Stokes Raman scattering}.
\newblock {\em J. Chem. Phys.}, 147(19), 2017.

\bibitem{Kamga1980}
Francois~M. Kamga and Mark~G. Sceats.
\newblock {Pulse-sequenced coherent anti-Stokes Raman scattering spectroscopy:
  a method for suppression of the nonresonant background}.
\newblock {\em Opt. Lett.}, 5(3):126, mar 1980.

\bibitem{Volkmer2002}
Andreas Volkmer, Lewis~D. Book, and X.~Sunney Xie.
\newblock {Time-resolved coherent anti-Stokes Raman scattering microscopy:
  Imaging based on Raman free induction decay}.
\newblock {\em Appl. Phys. Lett.}, 80(9):1505--1507, mar 2002.

\bibitem{Wurpel2002}
George W~H Wurpel, Juleon~M Schins, and Michiel M{\"{u}}ller.
\newblock {Chemical specificity in three-dimensional imaging with multiplex
  coherent anti-Stokes Raman scattering microscopy.}
\newblock {\em Opt. Lett.}, 27(13):1093--5, 2002.

\bibitem{Oron2002}
Dan Oron, Nirit Dudovich, and Yaron Silberberg.
\newblock {Single-Pulse Phase-Contrast Nonlinear Raman Spectroscopy}.
\newblock {\em Phys. Rev. Lett.}, 89(27):1--4, 2002.

\bibitem{Potma2006}
Eric~O Potma, Conor~L Evans, and X~Sunney Xie.
\newblock {Heterodyne coherent anti-Stokes Raman scattering (CARS) imaging}.
\newblock {\em Opt. Lett.}, 31(2):241, jan 2006.

\bibitem{Lotem1983}
Haim Lotem.
\newblock {Frequency modulation coherent anti-Stokes Raman spectroscopy
  (FM-CARS): A novel sensitive nonlinear optical method}.
\newblock {\em J. Chem. Phys.}, 79(5):2177--2180, 1983.

\bibitem{Ganikhanov2006b}
Feruz Ganikhanov, Conor~L. Evans, Brian~G. Saar, and X.~Sunney Xie.
\newblock {High-sensitivity vibrational imaging with frequency modulation
  coherent anti-Stokes Raman scattering (FM CARS) microscopy}.
\newblock {\em Opt. Lett.}, 31(12):1872, 2006.

\bibitem{Chen2010}
Bi-Chang Chen, Jiha Sung, and Sang-Hyun Lim.
\newblock {Chemical Imaging with Frequency Modulation Coherent Anti-Stokes
  Raman Scattering Microscopy at the Vibrational Fingerprint Region}.
\newblock {\em J. Phys. Chem. B}, 114(50):16871--16880, dec 2010.

\bibitem{Rocha-Mendoza2009}
Israel Rocha-Mendoza, Wolfgang Langbein, Peter Watson, and Paola Borri.
\newblock {Differential coherent anti-Stokes Raman scattering microscopy with
  linearly chirped femtosecond laser pulses}.
\newblock {\em Opt. Lett.}, 34(15):2258, aug 2009.

\bibitem{Chen2010a}
Bi-Chang Chen, Jiha Sung, and Sang-Hyun Lim.
\newblock {Frequency modulation coherent anti-Stokes Raman scattering (FM-CARS)
  microscopy based on spectral focusing of chirped laser pulses}.
\newblock {\em Multiphot. Microsc. Biomed. Sci. X}, 7569:756909, 2010.

\bibitem{Brinkmann2019}
Maximilian Brinkmann, Alexander Fast, Tim Hellwig, Isaac Pence, Conor~L. Evans,
  and Carsten Fallnich.
\newblock {Portable all-fiber dual-output widely tunable light source for
  coherent Raman imaging}.
\newblock {\em Biomed. Opt. Express}, 10(9):4437, sep 2019.

\bibitem{Yamashita2006}
Shinji Yamashita and Masahiro Asano.
\newblock {Wide and fast wavelength-tunable mode-locked fiber laser based on
  dispersion tuning}.
\newblock {\em Opt. Express}, 14(20):9399, 2006.

\bibitem{Brinkmann2016}
Maximilian Brinkmann, Sarah Janfr{\"{u}}chte, Tim Hellwig, Sven Dobner, and
  Carsten Fallnich.
\newblock {Electronically and rapidly tunable fiber-integrable optical
  parametric oscillator for nonlinear microscopy}.
\newblock {\em Opt. Lett.}, 41(10):2193, may 2016.

\bibitem{Gottschall2015}
Thomas Gottschall, Tobias Meyer, Martin Baumgartl, Cesar Jauregui, Michael
  Schmitt, J{\"{u}}rgen Popp, Jens Limpert, and Andreas T{\"{u}}nnermann.
\newblock {Fiber-based light sources for biomedical applications of coherent
  anti-Stokes Raman scattering microscopy}.
\newblock {\em Laser Photon. Rev.}, 9(5):435--451, sep 2015.

\bibitem{Sarri2019}
Barbara Sarri, Rafa{\"{e}}l Canonge, Xavier Audier, Emma Simon, Julien Wojak,
  Fabrice Caillol, C{\'{e}}cile Cador, Didier Marguet, Flora Poizat, Marc
  Giovannini, and Herv{\'{e}} Rigneault.
\newblock {Fast stimulated Raman and second harmonic generation imaging for
  intraoperative gastro-intestinal cancer detection}.
\newblock {\em Sci. Rep.}, 9(1):1--10, 2019.

\end{thebibliography}


\begin{thebibliography}{10}
\newcommand{\enquote}[1]{``#1''}

\bibitem{Ganikhanov2006}
F.~Ganikhanov, S.~Carrasco, X.~{Sunney Xie}, M.~Katz, W.~Seitz, and D.~Kopf,
  \enquote{{Broadly tunable dual-wavelength light source for coherent
  anti-Stokes Raman scattering microscopy},} {\protect\JournalTitle{Opt.
  Lett.}} \textbf{31}, 1292 (2006).

\bibitem{Jurna2006}
M.~Jurna, J.~P. Korterik, H.~L. Offerhaus, and C.~Otto, \enquote{{Noncritical
  phase-matched lithium triborate optical parametric oscillator for high
  resolution coherent anti-Stokes Raman scattering spectroscopy and
  microscopy},} {\protect\JournalTitle{Appl. Phys. Lett.}} \textbf{89}, 1--4
  (2006).

\bibitem{Tanimura2006}
T.~Tanimura, D.~Akamatsu, Y.~Yokoi, A.~Furusawa, and M.~Kozuma,
  \enquote{{Generation of a squeezed vacuum resonant on a rubidium D1 line with
  periodically poled KTiOPO4},} {\protect\JournalTitle{Opt. Lett.}}
  \textbf{31}, 2344 (2006).

\bibitem{Leindecker2011}
N.~Leindecker, A.~Marandi, R.~L. Byer, and K.~L. Vodopyanov,
  \enquote{{Broadband degenerate OPO for mid-infrared frequency comb
  generation},} {\protect\JournalTitle{Opt. Express}} \textbf{19}, 6296 (2011).

\bibitem{Klein2001}
M.~E. Klein, A.~Robertson, M.~A. Tremont, R.~Wallenstein, and K.~J. Boller,
  \enquote{{Rapid infrared wavelength access with a picosecond PPLN OPO
  synchronously pumped by a mode-locked diode laser},}
  {\protect\JournalTitle{Appl. Phys. B Lasers Opt.}} \textbf{73}, 1--10 (2001).

\bibitem{Yamashita2006}
S.~Yamashita and M.~Asano, \enquote{{Wide and fast wavelength-tunable
  mode-locked fiber laser based on dispersion tuning},}
  {\protect\JournalTitle{Opt. Express}} \textbf{14}, 9399 (2006).

\bibitem{Brinkmann2016}
M.~Brinkmann, S.~Janfr{\"{u}}chte, T.~Hellwig, S.~Dobner, and C.~Fallnich,
  \enquote{{Electronically and rapidly tunable fiber-integrable optical
  parametric oscillator for nonlinear microscopy},} {\protect\JournalTitle{Opt.
  Lett.}} \textbf{41}, 2193 (2016).

\bibitem{VanDriel1995}
H.~M. van Driel, \enquote{{Synchronously pumped optical parametric
  oscillators},} {\protect\JournalTitle{Appl. Phys. B}} \textbf{60}, 411--420
  (1995).

\bibitem{Kippenberg2004}
T.~J. Kippenberg, S.~M. Spillane, and K.~J. Vahala, \enquote{{Kerr-Nonlinearity
  Optical Parametric Oscillation in an Ultrahigh-Q Toroid Microcavity},}
  {\protect\JournalTitle{Phys. Rev. Lett.}} \textbf{93}, 083904 (2004).

\bibitem{Stern2018}
B.~Stern, X.~Ji, Y.~Okawachi, A.~L. Gaeta, and M.~Lipson,
  \enquote{{Battery-operated integrated frequency comb generator},}
  {\protect\JournalTitle{Nature}} \textbf{562}, 401--405 (2018).

\bibitem{Gaeta2019}
A.~L. Gaeta, M.~Lipson, and T.~J. Kippenberg, \enquote{{Photonic-chip-based
  frequency combs},} {\protect\JournalTitle{Nat. Photonics}} \textbf{13},
  158--169 (2019).

\bibitem{Levy2012}
J.~S. Levy, K.~Saha, Y.~Okawachi, M.~A. Foster, A.~L. Gaeta, and M.~Lipson,
  \enquote{{High-Performance Silicon-Nitride-Based Multiple-Wavelength
  Source},} {\protect\JournalTitle{IEEE Photonics Technol. Lett.}} \textbf{24},
  1375--1377 (2012).

\bibitem{Trocha2018}
P.~Trocha, M.~Karpov, D.~Ganin, M.~H.~P. Pfeiffer, A.~Kordts, S.~Wolf,
  J.~Krockenberger, P.~Marin-Palomo, C.~Weimann, S.~Randel, W.~Freude, T.~J.
  Kippenberg, and C.~Koos, \enquote{{Ultrafast optical ranging using
  microresonator soliton frequency combs},} {\protect\JournalTitle{Science
  (80-. ).}} \textbf{359}, 887--891 (2018).

\bibitem{Dutt2018}
A.~Dutt, C.~Joshi, X.~Ji, J.~Cardenas, Y.~Okawachi, K.~Luke, A.~L. Gaeta, and
  M.~Lipson, \enquote{{On-chip dual-comb source for spectroscopy},}
  {\protect\JournalTitle{Sci. Adv.}} \textbf{4}, e1701858 (2018).

\bibitem{Suh2016}
M.-G. Suh, Q.-F. Yang, K.~Y. Yang, X.~Yi, and K.~J. Vahala,
  \enquote{{Microresonator soliton dual-comb spectroscopy},}
  {\protect\JournalTitle{Science (80-. ).}} \textbf{354}, 600--603 (2016).

\bibitem{Papp2014}
S.~B. Papp, K.~Beha, P.~Del'Haye, F.~Quinlan, H.~Lee, K.~J. Vahala, and S.~A.
  Diddams, \enquote{{Microresonator frequency comb optical clock},}
  {\protect\JournalTitle{Optica}} \textbf{1}, 10 (2014).

\bibitem{Sharping2008}
J.~E. Sharping, \enquote{{Microstructure Fiber Based Optical Parametric
  Oscillators},} {\protect\JournalTitle{J. Light. Technol.}} \textbf{26},
  2184--2191 (2008).

\bibitem{Yang2018}
K.~Yang, S.~Zheng, Y.~Wu, P.~Ye, K.~Huang, Q.~Hao, and H.~Zeng,
  \enquote{{Low-repetition-rate all-fiber integrated optical parametric
  oscillator for coherent anti-Stokes Raman spectroscopy},}
  {\protect\JournalTitle{Opt. Express}} \textbf{26}, 17519 (2018).

\bibitem{Brinkmann2019a}
M.~Brinkmann, A.~Fast, T.~Hellwig, I.~Pence, C.~L. Evans, and C.~Fallnich,
  \enquote{{Portable all-fiber dual-output widely tunable light source for
  coherent Raman imaging},} {\protect\JournalTitle{Biomed. Opt. Express}}
  \textbf{10}, 4437 (2019).

\bibitem{DeVos2007}
K.~{De Vos}, I.~Bartolozzi, E.~Schacht, P.~Bienstman, and R.~Baets,
  \enquote{{Silicon-on-Insulator microring resonator for sensitive and
  label-free biosensing},} {\protect\JournalTitle{Opt. Express}} \textbf{15},
  7610 (2007).

\bibitem{Robinson2008a}
J.~T. Robinson, L.~Chen, and M.~Lipson, \enquote{{On-chip gas detection in
  silicon optical microcavities},} {\protect\JournalTitle{Opt. Express}}
  \textbf{16}, 4296 (2008).

\bibitem{Dhakal2014}
A.~Dhakal, A.~Z. Subramanian, P.~Wuytens, F.~Peyskens, N.~L. Thomas, and
  R.~Baets, \enquote{{Evanescent excitation and collection of spontaneous Raman
  spectra using silicon nitride nanophotonic waveguides},}
  {\protect\JournalTitle{Opt. Lett.}} \textbf{39}, 4025 (2014).

\bibitem{Kuyken2013}
B.~Kuyken, X.~Liu, R.~M. Osgood, R.~Baets, G.~Roelkens, and W.~M.~J. Green,
  \enquote{{A silicon-based widely tunable short-wave infrared optical
  parametric oscillator},} {\protect\JournalTitle{Opt. Express}} \textbf{21},
  5931 (2013).

\bibitem{Wang2015i}
K.-Y. Wang, M.~A. Foster, and A.~C. Foster, \enquote{{Wavelength-agile near-IR
  optical parametric oscillator using a deposited silicon waveguide},}
  {\protect\JournalTitle{Opt. Express}} \textbf{23}, 15431 (2015).

\bibitem{Liu2011}
X.~Liu, J.~B. Driscoll, J.~I. Dadap, R.~M. Osgood, S.~Assefa, Y.~A. Vlasov, and
  W.~M.~J. Green, \enquote{{Self-phase modulation and nonlinear loss in silicon
  nanophotonic wires near the mid-infrared two-photon absorption edge},}
  {\protect\JournalTitle{Opt. Express}} \textbf{19}, 7778 (2011).

\bibitem{Epping2015a}
J.~P. Epping, M.~Hoekman, R.~Mateman, A.~Leinse, R.~G. Heideman, A.~van Rees,
  P.~J. van~der Slot, C.~J. Lee, and K.-J. Boller, \enquote{{High confinement,
  high yield Si{\_}3N{\_}4 waveguides for nonlinear optical applications},}
  {\protect\JournalTitle{Opt. Express}} \textbf{23}, 642 (2015).

\bibitem{Roeloffzen2018}
C.~G.~H. Roeloffzen, M.~Hoekman, E.~J. Klein, L.~S. Wevers, R.~B. Timens,
  D.~Marchenko, D.~Geskus, R.~Dekker, A.~Alippi, R.~Grootjans, A.~van Rees,
  R.~M. Oldenbeuving, J.~P. Epping, R.~G. Heideman, K.~Worhoff, A.~Leinse,
  D.~Geuzebroek, E.~Schreuder, P.~W.~L. van Dijk, I.~Visscher, C.~Taddei,
  Y.~Fan, C.~Taballione, Y.~Liu, D.~Marpaung, L.~Zhuang, M.~Benelajla, and
  K.-J. Boller, \enquote{{Low-Loss Si3N4 TriPleX Optical Waveguides: Technology
  and Applications Overview},} {\protect\JournalTitle{IEEE J. Sel. Top. Quantum
  Electron.}} \textbf{24}, 1--21 (2018).

\bibitem{Agrawal2013a}
G.~P. Agrawal, \emph{{Nonlinear Fiber Optics}} (Academic Press, 2013), 5th ed.

\bibitem{Brinkmann2015}
M.~Brinkmann, S.~Dobner, and C.~Fallnich, \enquote{{Light source for narrow and
  broadband coherent Raman scattering microspectroscopy},}
  {\protect\JournalTitle{Opt. Lett.}} \textbf{40}, 5447 (2015).

\bibitem{Lupken2021a}
N.~M. L{\"{u}}pken, T.~W{\"{u}}rthwein, K.-J. Boller, and C.~Fallnich,
  \enquote{{Optical parametric amplification in silicon nitride waveguides for
  coherent Raman imaging},} {\protect\JournalTitle{Opt. Express}} \textbf{29},
  10424 (2021).

\bibitem{Lupken2020a}
N.~M. L{\"{u}}pken, T.~W{\"{u}}rthwein, J.~P. Epping, K.-J. Boller, and
  C.~Fallnich, \enquote{{Spontaneous four-wave mixing in silicon nitride
  waveguides for broadband coherent anti-Stokes Raman scattering
  spectroscopy},} {\protect\JournalTitle{Opt. Lett.}} \textbf{45}, 3873 (2020).

\bibitem{Poletti2008}
F.~Poletti and P.~Horak, \enquote{{Description of ultrashort pulse propagation
  in multimode optical fibers},} {\protect\JournalTitle{J. Opt. Soc. Am. B}}
  \textbf{25}, 1645 (2008).

\bibitem{Soong2011}
K.~Soong, R.~Byer, C.~McGuinness, E.~Peralta, E.~Colby, and R.~England,
  \enquote{{Experimental determination of damage threshold characteristics of
  IR compatible optical materials},} {\protect\JournalTitle{2011 Part. Accel.
  Conf. Proc.}} \textbf{277}, 277--279 (2011).

\bibitem{Kowligy2018}
A.~S. Kowligy, D.~D. Hickstein, A.~Lind, D.~R. Carlson, H.~Timmers, N.~Nader,
  D.~L. Maser, D.~Westly, K.~Srinivasan, S.~B. Papp, and S.~A. Diddams,
  \enquote{{Tunable mid-infrared generation via wide-band four-wave mixing in
  silicon nitride waveguides},} {\protect\JournalTitle{Opt. Lett.}}
  \textbf{43}, 4220 (2018).

\bibitem{Lefrancois2012}
S.~Lefrancois, D.~Fu, G.~R. Holtom, L.~Kong, W.~J. Wadsworth, P.~Schneider,
  R.~Herda, A.~Zach, X.~{Sunney Xie}, and F.~W. Wise, \enquote{{Fiber four-wave
  mixing source for coherent anti-Stokes Raman scattering microscopy},}
  {\protect\JournalTitle{Opt. Lett.}} \textbf{37}, 1652 (2012).

\bibitem{Rao2020}
A.~Rao, G.~Moille, X.~Lu, D.~A. Westly, D.~Sacchetto, M.~Geiselmann, M.~Zervas,
  S.~B. Papp, J.~Bowers, and K.~Srinivasan, \enquote{{Integrated photonic
  interposers for processing octave-spanning microresonator frequency combs},}
  {\protect\JournalTitle{ArXiv}}  (2020).

\bibitem{Tien2010}
M.-C. Tien, J.~F. Bauters, M.~J.~R. Heck, D.~J. Blumenthal, and J.~E. Bowers,
  \enquote{{Ultra-low loss Si{\_}3N{\_}4 waveguides with low nonlinearity and
  high power handling capability},} {\protect\JournalTitle{Opt. Express}}
  \textbf{18}, 23562 (2010).

\bibitem{Xue2016}
X.~Xue, Y.~Xuan, C.~Wang, P.-H. Wang, Y.~Liu, B.~Niu, D.~E. Leaird, M.~Qi, and
  A.~M. Weiner, \enquote{{Thermal tuning of Kerr frequency combs in silicon
  nitride microring resonators},} {\protect\JournalTitle{Opt. Express}}
  \textbf{24}, 687 (2016).

\bibitem{Bauters2011a}
J.~F. Bauters, M.~J.~R. Heck, D.~D. John, J.~S. Barton, C.~M. Bruinink,
  A.~Leinse, R.~G. Heideman, D.~J. Blumenthal, and J.~E. Bowers,
  \enquote{{Planar waveguides with less than 0.1 dB/m propagation loss
  fabricated with wafer bonding},} {\protect\JournalTitle{Opt. Express}}
  \textbf{19}, 24090 (2011).

\bibitem{Puckett2021}
M.~W. Puckett, K.~Liu, N.~Chauhan, Q.~Zhao, N.~Jin, H.~Cheng, J.~Wu, R.~O.
  Behunin, P.~T. Rakich, K.~D. Nelson, and D.~J. Blumenthal, \enquote{{422
  Million intrinsic quality factor planar integrated all-waveguide resonator
  with sub-MHz linewidth},} {\protect\JournalTitle{Nat. Commun.}} \textbf{12},
  934 (2021).

\end{thebibliography}

\end{document}